\def\Black{}
\def\Blue{}
\def\Brown{}
 \documentstyle[aps,epsfig,prl]{revtex}

\begin{document}

 \begin{titlepage}
 \null \noindent
 BARI-TH/00-388\\
 CERN-TH/2000-185\\
 DSF-2000/22\\
 UGVA-DPT/2000-06-1085\\

 \vspace{2cm}

 \begin{center}
 \Brown
 \Large\bf
 $B$ and $B_s$ decays into three pseudoscalar mesons \\
 and the determination of the angle $\gamma$ of the unitarity triangle
 \Black
 \end{center}
 \vspace{1.5cm}

 \begin{center}
 \begin{large}
 A. Deandrea$^a$, R. Gatto$^{b}$, M. Ladisa$^{c}$ G.
 Nardulli$^{c}$ and P. Santorelli$^d$\\
 \end{large}
 \vspace{0.7cm}
 $^a$ Theory Division, CERN, CH-1211 Gen\`eve 23, Switzerland\\
 $^b$ D\'epartement de Physique Th\'eorique, Universit\'e de Gen\`eve,\\
 24 quai E.-Ansermet, CH-1211 Gen\`eve 4, Switzerland\\
 $^c$ Dipartimento di Fisica, Universit\`a di Bari and INFN Bari,\\
 via Amendola 173, I-70126 Bari, Italy\\
 $^d$ Dipartimento di Scienze Fisiche, Universit\`a  di Napoli ``Federico
II''\\
 and INFN Napoli, Via Cintia, I-80126 Napoli, Italy\\
 \end{center}

 \vspace{1.5cm}

 \begin{center}
 \begin{large}
 \Brown
 {\bf Abstract}\\[0.5cm] \Black
 \end{large}
 \parbox{14cm}{
 We reconsider two classical proposals for the determination
 of the  angle $\gamma$ of the unitarity triangle:
 $B^\pm\to\chi_{c0}\pi^\pm\to\pi^+\pi^-\pi^\pm$ and $B_s\to\rho^0
 K_S\to\pi^+\pi^- K_S$. We point out the relevance, in both cases, of non
 resonant amplitudes, where the $\pi^+\pi^-$ pair is produced by weak decay
 of a $B^*$ ($J^P=1^-$) or $B_0$ ($J^P=0^+$) off-shell meson. In
 particular, for the $B$ decay channel, the inclusion of the $B_0$ pole
 completes some previous analyses and confirms their conclusions, provided a
 suitable cut in the Dalitz plot is performed; for the $B_s$ decay the
 inclusion of the $B^*,~B_0$ amplitudes enhances the role of the tree
 diagrams as compared to penguin amplitudes, which makes the theoretical
 uncertainty related to the $B_s\to\rho^0 K_S$ decay process less
 significant. While the first method is affected by theoretical
 uncertainties, the second one is cleaner, but its usefulness will depend on
 the available number of events to perform the analysis.}
 \end{center}

 \vspace{2cm}
 \noindent
 \Blue
 \Black

 \noindent
 July 2000
 \end{titlepage}

 \setcounter{page}{1}

 \preprint{CERN-TH/2000-185\\
 DSF-2000/22\\
 UGVA-DPT/2000-06-1085
 }
 \newcommand{\dd}{\displaystyle}
 \newcommand{\nn}{\nonumber}
 \newcommand{\be}{\begin{equation}}
 \newcommand{\ee}{\end{equation}}
 \newcommand{\bea}{\begin{eqnarray}}
 \newcommand{\eea}{\end{eqnarray}}
 \title{$B$ and $B_s$ decays into three pseudoscalar mesons and the
 determination of the angle $\gamma$ of the unitarity triangle}
 \author{A. Deandrea$^a$, R. Gatto$^{b}$, M. Ladisa$^{a,c}$ G.
 Nardulli$^{a,c}$ and P. Santorelli$^d$}
 \address{$^a$ Theory Division, CERN, CH-1211 Gen\`eve 23, Switzerland}
 \address{$^b$ D\'epartement de Physique Th\'eorique, Universit\'e de Gen\`eve,
 24 quai E.-Ansermet, CH-1211 Gen\`eve 4, Switzerland}
 \address{$^c$ Dipartimento di Fisica, Universit\`a di Bari and INFN Bari,
 via Amendola 173, I-70126 Bari, Italy}
 \address{$^d$ Dipartimento di Scienze Fisiche, Universit\`a  di Napoli
``Federico II''\\
 and INFN Napoli, Via Cintia, I-80126 Napoli, Italy}
 \date{June 2000}
 \maketitle

 \begin{abstract}
 We reconsider two classical proposals for the determination
 of the  angle $\gamma$ of the unitarity triangle:
 $B^\pm\to\chi_{c0}\pi^\pm\to\pi^+\pi^-\pi^\pm$ and $B_s\to\rho^0
 K_S\to\pi^+\pi^- K_S$. We point out the relevance, in both cases, of non
 resonant amplitudes, where the $\pi^+\pi^-$ pair is produced by weak decay
 of a $B^*$ ($J^P=1^-$) or $B_0$ ($J^P=0^+$) off-shell meson. In
 particular, for the $B$ decay channel, the inclusion of the $B_0$ pole
 completes some previous analyses and confirms their conclusions, provided a
 suitable cut in the Dalitz plot is performed; for the $B_s$ decay the
 inclusion of the $B^*,~B_0$ amplitudes enhances the role of the tree
 diagrams as compared to penguin amplitudes, which makes the theoretical
 uncertainty related to the $B_s\to\rho^0 K_S$ decay process less
 significant. While the first method is affected by theoretical
 uncertainties, the second one is cleaner, but its usefulness will depend on
 the available number of events to perform the analysis.
  \end{abstract}

 \section{Introduction}
 In the next few years dedicated $e^+ e^-$ machines at Cornell,
 SLAC and KEK and hadronic machines such as LHC will explore in
 depth several aspects of CP violations in the realm of
 $B$-physics. In particular the three angles $\alpha,\beta$ and
 $\gamma$ of the unitarity triangle will be extensively studied not
 only to nail down the Cabibbo-Kobayashi-Maskawa (CKM) matrix and
 its encoded mechanism for CP violations, but also to examine the
 possibility of deviations from the pattern expected in the
 Standard Model. Some analyses, based on  combined CDF and Aleph
 data \cite{CDF,Aleph} on $\sin\ 2\beta$, $\sin 2\beta\ =\ 0.82 ^+_-
 0.39$, as well as on CLEO results
 \cite{CLEO} and other constraints on the unitarity triangle, have
 been already used in \cite{parodi} to get limits on the three
 angles $\alpha, \, \beta$ and $ \gamma$. Although preliminary and
 based on a number of theoretical inputs, these results are worth
 to be quoted, as they represent theoretical and phenomenological
 expectations to be confirmed or falsified by the experiments to
 come\footnote{The fitted value of $\sin 2\beta$,
 which corresponds to the value (\ref{beta}), is
 $\sin 2\beta=0.750^{+0.058}_{-0.064}$ \cite{parodi}~.}:
 \bea
 && \beta\ =24.3^0~{\rm or~} 65.7^0
 \label{beta}\\
 &&\gamma\ =55.5^{+6.0}_{-8.5}
 \label{gamma}\\
 &&\alpha\ = \pi - \beta - \gamma ~.
 \label{alpha}
 \eea
 The first
 angle to be measured with a reasonable accuracy will be $\beta $,
 by the study of the channel $B\to J /\psi K_S$, which is free from
 the theoretical uncertainties related to the evaluation of
 hadronic matrix elements of the weak  Hamiltonian. A few
 strategies for the determination of $\alpha$ have been also
 proposed, most notably those based on the study of the channels
 $B\to\pi\pi$ and $B\to\rho\pi$ \cite{gronau}, \cite{snyder}. For
 this last channel a recent analysis \cite{deandrea} has stressed
 the role of non--resonant diagrams where one pseudoscalar meson is
 emitted by the initial $B$ meson with production of a $B^*$ or a
 positive parity $B_0$ ($J^P=0^+$) virtual state followed by the
 weak decay of these states into a pair of light pseudoscalar mesons.

 One of these diagrams (the virtual $B^*$ graph) has been examined
 also by other authors in the context of the determination of
 $\gamma$ \cite{deshpande},\cite{eilam},\cite{pham}. It is useful
 to point out that $\gamma$ appears at present the most
 difficult parameter of the unitarity triangle. In the
 recent years several methods have been proposed to measure this
 angle; some of them are theoretically clean, as they are based on
 the analysis of pure tree diagrams at quark level, such as $\bar b
 \to \bar u c \bar s$ and $\bar b \to \bar c u \bar s$ transitions.
 One of the benchmark modes  was proposed in \cite{gamma} and
 employs the decays $B^+ \to D^0 K^+$, $B^+ \to \overline{D^0} K^+$
 and $B^+ \to D_{\pm}^0 K^+$, where $D_{\pm}^0$ denotes
 CP eigenstates of the neutral $D$ meson system with CP
 eigenvalues $\pm 1$. The difference of the weak phases between the
 $B^+ \to D^0 K^+$ and the $B^+ \to  \overline{D^0} K^+$ amplitudes
 is $2 \gamma$, which would allow to extract the
 angle $\gamma$ by drawing two triangles with a common side: one of
 the triangles has  sides equal to $A(B^+ \to D^0 K^+)$, $A(B^+ \to
 \overline{D^0} K^+)$ and  ${\sqrt 2}A(B^+ \to D^0_+ K^+)$ respectively,
 and the other one has sides  $\dd A(B^- \to \overline{D^0}
 K^-)=e^{-2i\gamma}A(B^+ \to D^0 K^+) $, $A(B^- \to {D^0}
 K^-)=A(B^+ \to \overline{D^0} K^+)$ and  ${\sqrt 2}A(B^- \to D^0_+
 K^-)$. Even though this method is theoretically clean, it is
 affected by several experimental difficulties (for a discussion
 see \cite{ball}). One of these difficulties arises from the need
 to measure  the neutral $D$ meson decays into $CP$ eigenstates,
 but also  the other sides of the triangles present difficult
 experimental challenges. For example, if a hadronic decay (e.g.
 $D^0\to K^-\pi^+$) were used to tag the $D^0$ in the decay $B^+\to
 D^0 K^+$, there would be significant interference effects with the
 decay chain $B^+\to\overline{D^0}K^+\to K^-\pi^+K^+$ (through the
 doubly Cabibbo suppressed mode $\overline{D^0}\to K^-\pi^+$); if,
 on the other hand, the semileptonic channel $D^0\to\ell^+\nu_\ell
 X_s$ were used to tag the $D^0$, there would be contaminations
 from the background $B^+\to\ell^+\nu_\ell X_c$.

 The other benchmark modes for the determination of $\gamma$
 discussed in the recent review prepared for the Large Hadron
 Collider at CERN \cite{ball} have also their own experimental
 difficulties; for these reasons we consider worthwhile to consider
 other channels, already discussed in the past and somehow now
 disfavored because of their more intricate theoretical status. We
 are aware of these theoretical difficulties and it is the aim of
 the present paper to discuss them in some detail for two methods
 proposed for the determination of the angle $\gamma$. The first
 method was proposed in \cite{deshpande} and discussed subsequently
 by other authors \cite{eilam}, \cite{pham}: it is based  on the
 idea to analyze the charged $B$ CP-violating asymmetry, which
 arises from the interference between the resonant (at the invariant mass
 $m_{\chi_{c 0}}=3.417$ GeV) and non--resonant (the
 virtual $B^*$ graph) production of a
 pair of light pseudoscalar mesons in the decay $B\to 3$ light mesons.
 It is an aim of the present work to complete
 the analyses in \cite{deshpande,eilam,pham} by
 considering the channel $B \to 3 \pi$,
 including also the contribution of the virtual positive parity
 $B_0$ ($J^P=0^+$) state and the gluonic penguin operators. We
 shall therefore analyze the robustness of the conclusions in
 \cite{deshpande}, \cite{eilam} and \cite{pham} once these additional
 contributions are considered.

 The second analysis we consider here is the possible
 determination of $\gamma$ by means of the $B_s\to\rho^0 K_S$ decay
 mode. Also this process has been considered in the past
 \cite{BaBar}, but it is presently less emphasized
 because the tree level contribution, that one hopes to estimate
 more reliably, is suppressed by the smallness of the Wilson
 coefficient $a_1$. As we shall notice below, the non--resonant tree
 contributions to this decay (i.e. $B^*$ and $B_0$) are
 proportional to the large Wilson coefficient  $a_2$ ($a_2\approx
 1$); therefore we expect that their inclusion can reduce the
 theoretical uncertainties arising from the penguin terms.
 This channel could be a second generation experiment provided a sufficient
 number of events can be collected, once $x_s$, the
 mixing parameter for the $B_s - \bar B_s$ system, and $\beta$ have
 been determined by other experiments.

 \section{$B\to\chi_{c 0} \pi$ decays}
 We consider in this section the decay mode \be B^-
 \to \pi^+ \pi^- \pi^-~, \ee as well the CP-conjugate mode $B^+ \to
 \pi^- \pi^+ \pi^+ $, in the invariant mass range $m_{\pi^+
 \pi^-} \simeq m_{\chi_{c0}} \simeq 3.417$ GeV. For this decay mode
 we have both a resonant contribution coming from the decay $B^-
 \to \chi_{c0} \pi^- \to \pi^+ \pi^- \pi^-$ and several non
 resonant contributions. According to the analysis performed in
 \cite{deshpande}-\cite{pham}, this decay mode can be used to
 determine $\sin\gamma$ by looking for the charged $B$ asymmetry
 arising from two amplitudes: the resonant production {\it via}
 $\chi_{c0}$ decay and non--resonant amplitudes. Among the non
 resonant terms, we have included  the $B^*$ pole, which is the
 largest among the contributions considered in \cite{deshpande}
 \footnote{Other less important terms discussed in \cite{deshpande}
 include a long-distance type diagram, where an intermediate highly off-shell
 pion is exchanged among the incoming $B$ meson and the outgoing pions, and a
 short-distance diagram, where the outgoing pions are produced in a point-like
 effective interaction by the weak decay of the $B$ meson; we agree with the
 authors in \cite{deshpande} on  the
 smallness of these neglected terms.}. The authors in \cite{pham}
 have considered other decay modes in the same kinematical region,
 by analyzing the partial width asymmetry in $B^{\pm} \to M \bar M
 \pi^{\pm}$ decays ($M=\pi^+, K^+, \pi^0, \eta$). Spotting the
 decay mode $B^- \to \pi^+ \pi^- \pi^-$, they estimate an asymmetry
 given approximately by $0.33\, \sin
 \gamma$, which, however, seems to be  sensitive to the choice of
 the  parameters \cite{pham}.

 Our interest in this decay channel has been triggered
 by the study  of  a different invariant mass
 region (i.e. $m_{\pi \pi} \simeq m_\rho$)\cite{deandrea}, where
 also the contribution of the $B_0$ pole ($J^P = 0^+$, with an
 estimated mass 5.697 GeV) was found to be significant; therefore
 we include it in the present analysis, which represents an
 improvement in comparison to previous work. The second
 improvement we consider is the inclusion of the gluonic
 penguin operators. We refer to the paper \cite{deandrea} for a
 full discussion of the formalism and we list here only the
 relevant contributions $A_{\chi_{c0}}$, $A_{ B^{*}} $ and
 $A_{B_0}$ to the decay amplitude:
 \bea A_{\chi_{c0}} &=& K_{\chi}\ \left(\ \frac{1}{t -
 m_{\chi_{c0}}^2 + i\ m_{\chi_{c0}}\ \Gamma_{\chi_{c0}}} +
 \frac{1}{s - m_{\chi_{c0}}^2 + i\ m_{\chi_{c0}}\
 \Gamma_{\chi_{c0}}}\ \right) \; , \nn \\
 A_{B^{*}} &=&  K_{B^*}\ \left(\ \frac{\tilde \Pi(t,u)}{t -
 m_{B^*}^2 + i\ m_{B^*}\ \Gamma_{B^*}} + \frac{\tilde \Pi(s,
 u)}{s - m_{B^*}^2 + i\ m_{B^*}\ \Gamma_{B^*}}\ \right) \; , \nn \\
 A_{B_0} &=&  K_{B_0}\ (\ m_{B_0}^2 -
 m_\pi^2\ )\ \left(\ \frac{1}{t - m_{B_0}^2 + i\ m_{B_0}\
 \Gamma_{B_0}} + \frac{1}{s - m_{B_0}^2 + i\ m_{B_0}\
 \Gamma_{B_0}}\ \right)\; , \nn \\
 \label{amplitudes-chi}
 \eea
 where
 \bea
 &&u\  =\ (\ p_{\pi_1^-} + p_{\pi_2^-}\ )^2 \;\; ,\ s\
 =\ (\ p_{\pi^+} + p_{\pi_1^-}\ )^2 \;\; ,\ t\ =\ (\ p_{\pi^+} +
 p_{\pi_2^-}\ )^2 \; , \nn \\
 &&{\tilde \Pi} (x,y) = m^2_{\pi} -\frac{y}{2} +
 \frac{x ( M_B^2 -m_\pi^2 -x )}{4 m^2_{B^*}} \; .
 \label{kinematics-chi}
 \eea
 In (\ref{amplitudes-chi}) the values of the constants
 are:
 \bea
 K_{\chi} &=&  1.52\ 10^{-8}\ GeV^2\;\; ,
 \label{constants1-chi} \\
 K_{B^*} &=& -4\ \sqrt{2}\ g\
 m_B^2\ A_0^{B^* \pi}\ \frac{G_F}{\sqrt{2}}\ \left[\ V_{ub}\
 V_{ud}^*\ a_2 - V_{tb}\ V_{td}^*\ \left(\ a_4 - a_6\
 \frac{m_\pi^2}{m_q\ (\ m_b + m_q\ )}\ \right)\ \right] \; ,
 \label{constants2-chi} \\
 K_{B_0} &=&  h\ \sqrt{\frac{m_B}{m_{B_0}}}\ (\ m_{B_0}^2 - m_B^2\
 )\ F_0^{B \pi}\ \frac{G_F}{\sqrt{2}}\ \left[\ V_{ub}\ V_{ud}^*\
 a_2 - V_{tb}\ V_{td}^*\ \left(\ a_4 - a_6\ \frac{m_\pi^2}{m_q\ (\
 m_b + m_q\ )}\ \right)\ \right] \; .
 \label{constants3-chi}
 \eea
 The numerical value in
 (\ref{constants1-chi}) is derived in \cite{eilam}, where the
 resonance amplitude is given by
 \be
 R(s) = \alpha_1\ \alpha_2\ \frac{\sqrt{\Gamma_{\chi_{c0}}\
 m_{\chi_{c0}}}}{s - m_{\chi_{c0}}^2 +i\ \Gamma_{\chi_{c0}}\
 m_{\chi_{c0}}} \;\; . \label{B-W}
 \ee
 Normalizing the decay
 rate of $B^+ \to \chi_{c0} \pi^+ \to \pi^+ \pi^- \pi^+$ by the
 total $B$ decay rate, the product $\alpha_1\ \alpha_2$ in
 (\ref{B-W}) is given by the product of the corresponding branching
 ratios~:
 \be
 2\ \pi\ \alpha_1^2\ \alpha_2^2 = Br(B^+ \to \chi_{c0}) \times
 Br(\chi_{c0} \to \pi^+ \pi^-) \;\; . \label{K-chi} \ee In
 \cite{eilam} the product of the branching ratios in (\ref{K-chi})
 is estimated to be about $5 \times 10^{-7}$~, which gives  the
 numerical value  in (\ref{constants1-chi})~.

 As to the numerical values of the constants appearing in
 (\ref{constants2-chi}) and (\ref{constants3-chi}), we use the same
 values adopted in \cite{deandrea}~: $g=0.4$, $h=-0.54$,
 $m_{B^*}=m_B=5.28$ GeV, $m_{B_0}=5.697$ GeV, $\Gamma_{B_0}=0.36$
 GeV, $\Gamma_{B^*}=0.2$ KeV, $m_b=4.6$ GeV, $m_q \approx m_u
 \approx m_d \simeq 6$ MeV, $A_0^{B^* \pi}=0.16$, $F_0^{B
 \pi}=-0.19$. These numerical estimates agree with results obtained
 by different methods: QCD sum rules \cite{QCD-sum-rules},
 potential models \cite{pot-mod}, effective Lagrangian
 \cite{eff-lagr}, NJL-inspired models \cite{NJL}~. Moreover we use
 the following values of the Wilson coefficient~:
 $C_1=-0.226$, $C_2=1.1$, $C_3=0.012$, $C_4=-0.029$, $C_5=0.009$ and
 $C_6=-0.033$~, with $a_2=C_2+C_1/3$, $a_1=C_1+C_2/3$. The Wilson
 coefficients are obtained in the HV scheme \cite{buras},
 with $\Lambda^{(5)}_{\bar {MS}}=225$ MeV, $\mu = \bar m_b (m_b) =
 4.40$ GeV and $m_t = 170$ GeV. For the CKM mixing matrix
 \cite{CKM} we use the Wolfenstein parameterization
 \cite{wolfenstein}: $V_{ub}= A\ \lambda^3(\rho- i \eta)$,
 $V_{tb}=1$, $V_{ud}=1-\lambda^2/2$, $V_{td}=A\ \lambda^3\ (1-\rho
 -i\eta)$, $V_{cb}=A\ \lambda^2$, $V_{cs}=1-\lambda^2/2$ and
 $V_{ts}=-A\ \lambda^2$. We take $\lambda=0.22$ and $A=0.831$;
 moreover, since $\eta$ is better known than $\rho$ we take it at
 the value provided by the present analyses of the CKM matrix:
 $\eta=0.349$ \cite{parodi}~. It follows that $\rho$ will be given,
 in terms of $\gamma$, by $\dd \rho=\eta/\tan\gamma$.

 The asymmetry is given by
 \be
 {\cal A} = \frac{\Gamma (B^+ \to \pi^- \pi^+ \pi^+) - \Gamma (B^- \to \pi^+
 \pi^- \pi^-)}{\Gamma (B^+ \to \pi^- \pi^+ \pi^+) +
 \Gamma (B^- \to \pi^+ \pi^- \pi^-)} \;\; .
 \label{asym}
 \ee
 \par
 \noindent By introducing only the $\chi_{c0}$ and $B^*$
 contributions, we reproduce, within the theoretical uncertainties,
 the results of \cite{pham}. However the introduction of the $B_0$
 pole contribution dramatically reduces the asymmetry, because this
 contribution to the asymmetry is opposite to the $B^*$ term. We
 have observed that this cancellation arises from a change of sign
 around the $\chi_{c0}$ resonance and therefore we change a little
 bit the procedure by defining a cut in the Dalitz plot. We
 integrate in the region defined by
 \bea
 m_{\chi_{c0}} - 2\ \Gamma_{\chi_{c0}} &\leq& \sqrt{s} \leq m_{\chi_{c0}} + 2\
 \Gamma_{\chi_{c0}} \; , \nn \\
 m_{\chi_{c0}} &\leq&  \sqrt{t} \; ,
 \label{cut}
 \eea
 or
 \bea
 m_{\chi_{c0}} - 2\ \Gamma_{\chi_{c0}} &\leq& \sqrt{t} \leq
 m_{\chi_{c0}} + 2\ \Gamma_{\chi_{c0}} \; , \nn \\
 m_{\chi_{c0}}  &\leq& \sqrt{s} \; ,
 \label{cut-off}
 \eea
 where
 $\Gamma_{\chi_{c0}}=14$ MeV.
 It may be useful to observe that the integration over
 the whole available space in the Mandelstam plane around the $\chi_{c 0}$
 resonance gives $Br(B^-
 \to \pi^- \pi^- \pi^+) \simeq Br(B^+ \to \pi^- \pi^+ \pi^+) = 5.27
 \times 10^{-7}$ and therefore the cut-off procedure introduces a
 reduction of a factor 5 in the branching ratio.

\begin{figure}[ht]
 \epsfxsize=7cm
 \centerline{\epsffile{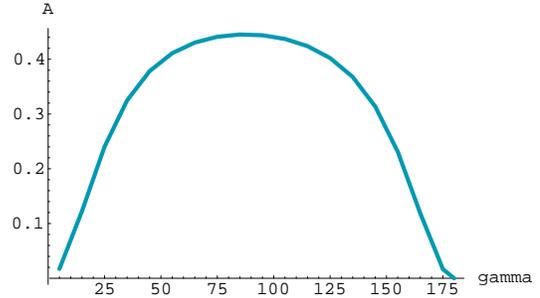}}
 \caption{Asymmetry as a function $\gamma$ for $B \to \chi_{c0} \pi$. }
 \label{f:asymchi}
 \end{figure}

 For the asymmetry we obtain the result in Fig. \ref{f:asymchi}.
 For $\gamma \simeq 55^o$, it can be approximated by
 ${\cal A}_{cut}=0.48\ \sin\gamma$.
 In order to assess the
 relevance of the $B_0$ pole, we report in Table \ref{t:tab1}
 the contribution to the branching and to the asymmetry of the different
 contributions for a particular value of $\sin\gamma$.

 We observe that the inclusion of the next low-lying
 state $B_0$ does not alter significantly the conclusions obtained
 in previous works, where basically only the $B^*$ non--resonant
 term was considered; however this conclusion can be obtained only if
 a convenient cut in the Dalitz plot
 is included. We also observe that the calculations performed in this
 section are not sensitive to the inclusion of the gluonic
 penguin contributions.

 To get an estimate of the dependence of our result on the parameters,
 we considered the following intervals for the couplings $g$ and $h$.
 For $h=-0.54$ and $g=0.4\pm 0.1$ we obtain (at $\sin \gamma = 0.82$)
 an asymmetry ${\cal A}_{cut}=0.41^{+0.05}_{-0.12}$; for $g=0.4$ and
 $h=-.54\pm 0.16$ we have an asymmetry
 ${\cal A}_{cut}=0.41^{+0.03}_{-0.04}$. The corresponding variation on
 $\gamma$ is extremely large ($30^o$ to $150^o$ degrees) because the
 asymmetry is rather flat in that region. We conclude that due to the
 theoretical uncertainties inherent to this method, the channel
 $\chi_{c0} \pi$ can hardly be useful for a precise determination of
 the angle $\gamma$.

 \section{$B_s\to\rho^0 K_S$ decay}
 In the decay $B_s\to\rho^0 K_S$
 the final state is a CP eigenstate; in this case one can
 measure either the time dependent asymmetry
 \be
 R_1(t)=\frac{\Gamma(B_s(t)\to \rho^0 K_S) \ - \ \Gamma(\bar B_s(t)\to \rho^0
 K_S)}
 {\Gamma(B_s(t)\to \rho^0 K_S) \ + \ \Gamma(\bar B_s(t)\to \rho^0
 K_S)}\ ,
 \ee
 or the time integrated ($t>0$) asymmetry:
 \be
 R_2=\frac{\int_0^\infty dt\ \left[\Gamma(B_s(t)\to \rho^0 K_S) \ - \
  \Gamma(\bar B_s(t)\to \rho^0 K_S)\right]}
 {\int_0^\infty dt\ \left[ \Gamma(B_s(t)\to \rho^0 K_S) \ + \ \Gamma(\bar
 B_s(t)\to \rho^0 K_S)\right]}\ .
 \ee
 Let us define
 \be
 x_s= \frac{\Delta m_s}{\Gamma} \label{x}
 \ee
 where $ \Delta m_s$ is the mass difference between the mass
 eigenstates and $\Gamma \approx \Gamma (B_s) \approx \Gamma (\bar B_s)$
 and
 \be
 A=A(B_s \to \rho^0 K_S),~~~~~~{\bar A}=A({\bar B_s} \to \rho^0 K_S)
 \ee
 \bea
 && A=|A_T|e^{i(\phi_T+\gamma) } +|A_P|e^{i(\phi_P-\beta) }  \\
 && \bar A=|A_T|e^{i(\phi_T-\gamma) } +|A_P|e^{i(\phi_P+\beta) } \; .
 \eea
 Here $\phi_T$ and  $\phi_P$ are strong phases of the tree and
 penguin amplitudes, $|A_T|$ and $|A_P|$ their absolute values and
 $\beta$ and $\gamma$ the weak phases of the $V^*_{td}$ and $V^*_{ub}$
 CKM matrix elements. The mixing between $B_s$ and $\bar B_s$,
 parameterized by the $x_s$ parameter in (\ref{x}) introduces no weak
 phase.

\begin{figure}[ht!]
 \epsfxsize=7cm
 \centerline{\epsffile{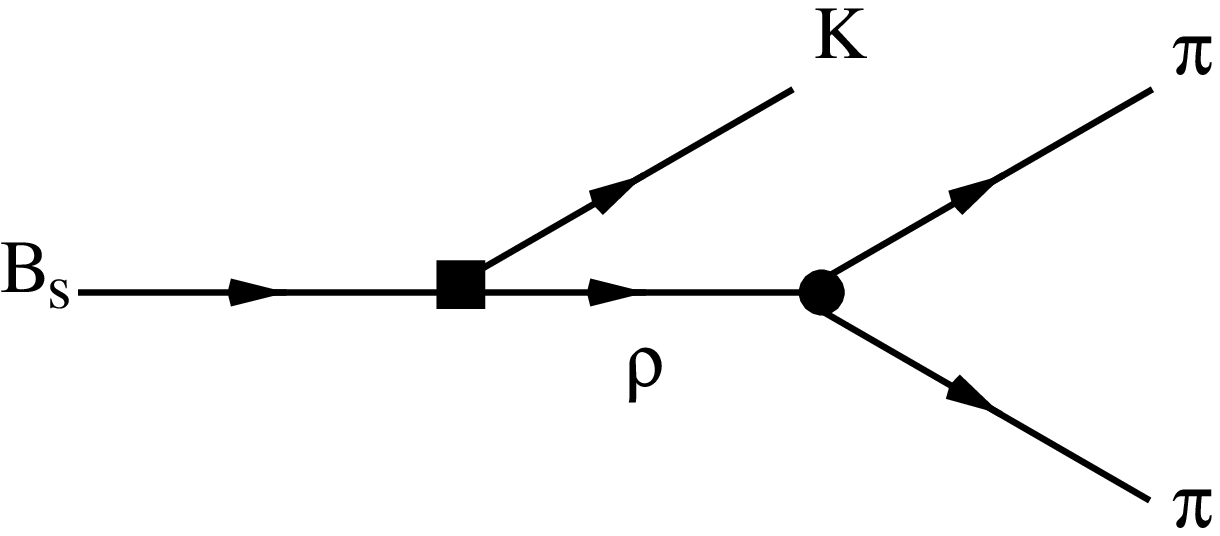}}
 \caption{The $\rho$ Feynman diagram for the
 $B_s \to K_S\ \pi^-\ \pi^+$
 decay. The circle and the box represent,
 respectively, the strong and the weak interaction vertex.}
 \label{f:diagram1}
 \end{figure}

\begin{figure}[ht!]
 \epsfxsize=7cm
 \centerline{\epsffile{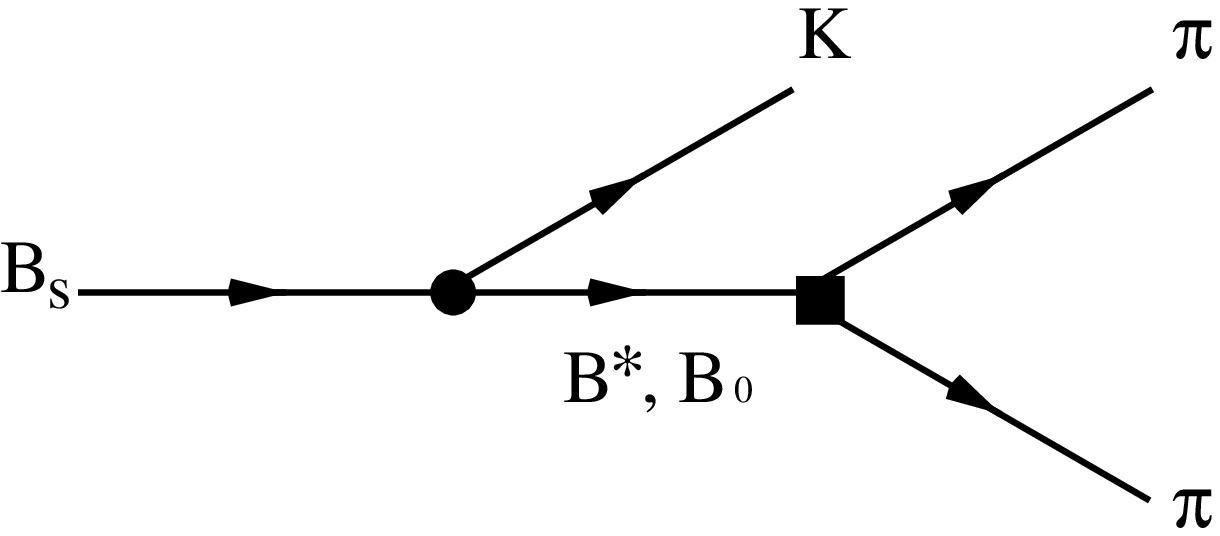}}
 \caption{The $B^*$, $B_0$ Feynman diagrams for the
 $B_s \to K_S\ \pi^-\ \pi^+$
 decay. The circle and the box represent,
 respectively, the strong and the weak interaction vertex.}
 \label{f:diagram2}
 \end{figure}

 Both the $\rho^0$ diagram (fig. \ref{f:diagram1}) and the
 $B^*,\ B_0$ non--resonant diagrams, with a cut in the $\pi^+,\ \pi^-$
 pair at $m_{\pi\pi}= m_\rho\pm 2\Gamma_\rho$ (fig. \ref{f:diagram2})
 contribute to $A_T$ and $A_P$, that are therefore given as follows
 \bea
 A_T&=&|A_T|e^{i(\phi_T+\gamma) }= A^T_\rho +A^T_{B^*}+A^T_{B_0} \; ,
 \label{a} \\
 A_P&=&|A_P|e^{i(\phi_P-\beta) }= A^P_\rho +A^P_{B^*}+A^P_{B_0} \; .
 \label{abar}
 \eea

 The amplitudes are computed in the factorization approximation from
 the weak non leptonic Hamiltonian as given by \cite{buras}; our approach
 is similar to the
 one employed in ref. \cite{deandrea} where a full description of the
 method is given. We get ($Q=T,P$):
 \bea
 A^Q_{\rho} &=& K^Q_{\rho}\
 \frac{t - t'}{u - m_\rho^2 + i\ m_\rho\ \Gamma_{\rho}} \; , \nn \\
 A^Q_{ B^{*}} &=& K^Q_{ B^{*}}\ \frac{t - t'}{u - m_{B^*}^2 + i\ m_{B^*}\
 \Gamma_{B^*}} \; , \nn \\
 A^Q_{B_0} &=& -\ K^Q_{B_0}\ \frac{m_{B_0}^2 - m_{B_s}^2}{u - m_{B_0}^2 + i\
 m_{B_0}\ \Gamma_{B_0}}\ \; , \nn \\
 \label{amplitudes}
 \eea
 where
 \be
 u\  =\ (\ p_{\pi^-} + p_{\pi^+}\ )^2 \;\;\; ,\
 t\  =\ (\ p_{K} + p_{\pi^-}\ )^2 \;\;\; ,\
 t'\ =\ (\ p_{K} + p_{\pi^+}\ )^2 \; .
 \label{kinematics}
 \ee
 In (\ref{amplitudes}) the values of the constants are~:
 \bea
 K^T_{\rho} &=& \frac{G_F}{2 \sqrt{2}}\
 \ V_{ud}^*\ V_{ub}\ a_1 \
 g_{\rho \pi \pi}\ f_{\rho}\ F_1^{B_s K} \;\; , \label{constants1} \\
 K^P_{\rho} &=& \frac{G_F}{2 \sqrt{2}}\
   V_{td}^*\ V_{tb}\ a_4\ \
 g_{\rho \pi \pi}\ f_{\rho}\ F_1^{B_s K} \;\; ,  \label{constants2}  \\
 K^T_{ B^*} &=& 4\ A_0^{ B^{*} \pi}\ \frac{G_F}{\sqrt{2}}\
 V_{ud}^*\ V_{ub}\ a_2  g\
 \frac{f_{\pi}}{f_{K}}\ m_{B_s}\ m_{B^*} \;\; , \label{constants3} \\
 K^P_{ B^*} &=& -\ 4\ A_0^{ B^{*} \pi}\ \frac{G_F}{\sqrt{2}}\
 V_{td}^*\ V_{tb}\ \left(\ a_4 -
 a_6\ \frac{m_\pi^2}{m_q\ (\ m_b + m_q\ )}\ \right)\ g\
 \frac{f_{\pi}}{f_{K}}\ m_{B_s}\ m_{B^*} \;\; , \label{constants4} \\
 K^T_{B_0} &=& \tilde F_0^{B_0 \pi}\ \frac{m_{B_0}^2 - m_\pi^2}{m_{B_0}}\
 \frac{G_F}{\sqrt{2}}\  V_{ud}^*\
 V_{ub}\ a_2  \sqrt{m_{B_0}\
 m_{B_s}}\ h\ \frac{f_{\pi}}{f_{K}} \;\; , \label{constants5} \\
 K^P_{B_0} &=& -\tilde F_0^{B_0 \pi}\ \frac{m_{B_0}^2 - m_\pi^2}{m_{B_0}}\
 \frac{G_F}{\sqrt{2}}\  V_{td}^*\ V_{tb}\ \left(\ a_4 - a_6\
\frac{m_\pi^2}{m_q\
 (\ m_b + m_q\ )}\ \right)\  \sqrt{m_{B_0}\
 m_{B_s}} \frac{h f_{\pi}}{f_{K}} \; ,
 \label{constants6}
 \eea
 where $g_{\rho \pi \pi}=5.8$, $f_\rho=0.15$ GeV$^2$ \cite{ball-frere-tytgat},
 $m_\rho=770$ MeV, $\Gamma_\rho=150$ MeV, $f_\pi=130$ MeV, $f_K=161$ MeV,
 $\tilde F_0^{B_0 \pi}=-0.19$, $F_1^{B_s K}=-0.19$,$m_{B_s}=5.37$ GeV
 \cite{deandrea}. From these equations the parameters appearing in
 (\ref{a}), (\ref{abar}) can be obtained.
 The time integrated asymmetry is
 \be
 {\mathcal{A}} = \frac{x_s \left[ \sin 2 \gamma -\alpha_1\sin 2 \beta-
 2 \alpha_2\sin(\beta-\gamma)\right]-2\alpha_3\sin(\gamma+\beta)}{
 (1+x_s^2)\left[1+\alpha_1+2\alpha_2
 \cos(\beta+\gamma)\right]} \; .
 \ee
 Numerically we obtain:
 \bea
 && \alpha_1= \frac{\int d \Omega|A_P|^2 }{\int d \Omega|A_T|^2}=0.06 \; ,
 \nn \\
 && \alpha_2= \frac{\int d \Omega\cos(\phi_T-\phi_P)|A_P A_T| }{\int d
 \Omega|A_T|^2}=-0.09 \; , \nn \\
 && \alpha_3= \frac{\int d \Omega \sin(\phi_T-\phi_P)|A_P A_T| }{\int d
 \Omega|A_T|^2}=0.015 \; .
 \label{alpha-1-2-3}
 \eea In these equations integrations are performed in a band
 around the $\rho$ mass: $m_\rho \pm 200$ MeV.

 For illustrative purposes we consider the value $x_s=23$, $\beta=65.7^0$
 and $\gamma=55.5^0$, corresponding to the central values in \cite{parodi};
 one obtains an asymmetry of $3.5 \%$
 \footnote{For the solution $\beta=24.3^0$ and the same values of $\gamma$
 and $x_s$ one gets for the asymmetry again $3.5 \%$ as the coefficients
 $\alpha_1$, $\alpha_2$, $\alpha_3$ are small and the asymmetry
 can roughly be approximated by $\sin 2\gamma /x_s$.}.

 It can be observed that the channel $B_s\to\rho^0 K_S$ has been discussed
 elsewhere in the literature \cite{fleischer}, but somehow discarded
 for two reasons. First the asymmetry contains a factor
 $x_s/(1+x_s^2)$ which, in view of the large mixing between $B_s$ and
 $\bar B_s$, is rather small. Second, as it is clear from eqs.
 (\ref{constants6}), the ratio of
 the penguin to the tree amplitudes can be
 large, if one includes only the $\rho^0$-resonant diagrams
 {\footnote{without the $B^*$ and $B_0$ contribution the parameters
 of eq. (\ref{alpha-1-2-3} would be larger $\alpha_1=0.26$,
 $\alpha_2=-0.27$, $\alpha_3=-0.45$}}; indeed the
 $\rho^0$ contribution is proportional to the Wilson coefficient $a_1$ which is
 small. As to the first point a small asymmetry can still be useful for
 determining $\gamma$ provided a sufficient number of events is available
 (see below); as to the second point
 the inclusion of the non--resonant contribution $B^*$, $B_0$ is of
 some help in this context, as the tree contribution is proportional to the
 Wilson coefficient $a_2 \simeq 1.0$ for these diagrams.

 A reliable estimate of the branching ratio is difficult (because
 of the uncertainty on the $a_1$ parameter). The effect on the asymmetry
 is to reduce the influence of the penguin operator in the final result
 as can be deduced from eq. (\ref{alpha-1-2-3}).
 In order to assess the validity of the method for the determination of the
 asymmetry, we varied the penguin
 contribution by varying the $\alpha_i$ parameters of eq. (\ref{alpha-1-2-3})
 by $50 \%$ \footnote{The reason could be a violation of
 factorization or a variation in the parameters used to estimate the penguin
 contribution.}. Our results for the asymmetry vary by $10 \%$
 (assuming $\gamma=55.5^o$) and the value of $\gamma$ that one can deduce
 is $55.5^{+3}_{-5}$ degrees due to this uncertainty.

\begin{figure}[ht!]
 \epsfxsize=7cm
 \centerline{\epsffile{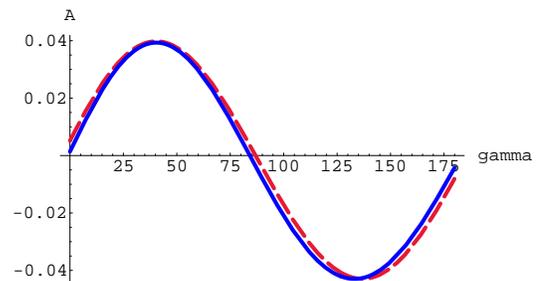}}
 \caption{The relevant asymmetry in the decay channel
 $B_s \to K_S\ \pi^+ \pi^-$ as a function of $\gamma$.
 The solid line corresponds to $\beta =24.3^o$ while the
 dashed one to $\beta =65.7^o$}
 \label{f:asym}
 \end{figure}

 In fig. (\ref{f:asym})
 we report the asymmetry as a function of the angle $\gamma$ (for $x_s=23$
 and two values of $\beta$).

 Let us conclude this analysis with a discussion on the reliability of the
 $B_s$ decay mode for the determination of $\gamma$. An estimate of
 the sensitivity of the method can be obtained by comparing it, as an
 example, to $B_s \to J/\Psi K_S$.
 The branching ratio for $B_s \to J/\Psi K_S$ is expected to be $2.0 \times
 10^{-5}$ \cite{ball}, while the branching $B_s \to \rho^0 K_S$ is roughly
 one order of magnitude smaller {\footnote{The precise value critically
 depends on the parameter $a_1$ which is the result of the partial
 cancellation of the Wilson coefficient $c_1$ and $c_2$ and on the
 validity of the factorization approximation. In \cite{deandrea2} an estimate
 of $(1 \pm 0.5) \times 10^{-6}$ is given; with the values adopted in the
 present paper we get $2 \times 10^{-7}$ because a much smaller value of $a_2$
 is used. Note however that the asymmetry is largely independent of the
 precise values of the parameters used to obtain the branching ratio.}}.
 The event yield for the $B_s \to J/\Psi K_S$ channel is estimated to be
 4100 event per year by a selection method developed by the CMS collaboration
 at the LHC (with a $p_T$ cut $> 1.5$ GeV/c on the pions from the $K_S$
 decays to suppress the combinatoric background). Assuming a similar
 selection method for $B_s \to \rho^0 K_S$, one could obtain $\simeq 410$
 events per year and $\simeq 2\times 10^3$ in 5 years, which would produce an
 uncertainty of $\pm 17^o$ on $\gamma$ (assuming $x_s = 23$ and
 $\gamma \simeq 55^o$) to be compared to the estimated
 error of $\pm 9^o$ degrees within 3 years at LHC for $B_s \to J/\Psi K_S$.
 Therefore even if the mode $B_s \to \rho^0 K_S$ is less competitive
 than the $B_s \to J/\Psi K_S$ one, it is not dramatically so if the
 branching ratio is not too small, and could be considered as a complementary
 analysis for the determination of $\gamma$. The final assessment
 of the feasibility will be clear as soon as an experimental determination
 of the branching ratio for $B_s \to \rho^0 K_S$ is available.

 \section{Conclusions}
 In this paper we have reviewed two classical methods proposed in the past
 few years for the determination of the angle $\gamma$:
 $B^\pm\to\chi_{c 0}\pi^\pm\to\pi^+\pi^-\pi^\pm$ and $B_s\to\rho^0
 K_S\to\pi^+\pi^- K_S$. For the first decay channel we have included,
 besides the $B^*$ non--resonant diagram, the $B_0$ ($J^P=0^+$) off--shell
 meson contribution. This calculation completes previous analyses and
 confirms their results, provided a suitable cut in the Dalitz plot is
 performed; however it appears that this method is subject to a large
 uncertainty on the determination of $\gamma$ coming from the allowed
 variation in the theoretical parameters because the asymmetry is rather
 flat in the region of interest. For the second
 channel we have pointed out the relevance of the two non--resonant
 amplitudes, i.e. the mechanism where the $\pi^+\pi^-$ pair
 is produced by weak decay of a $B^*$ ($J^P=1^-$) or $B_0$ ($J^P=0^+$)
 off--shell meson. The inclusion of these terms enhances the role of
 the tree diagrams as compared to penguin amplitudes, which makes the
 theoretical uncertainty related to the $B_s\to\rho^0 K_S$ decay process less
 significant. This method can be considered for a complementary
 analysis for the determination of $\gamma$, provided a sufficient
 number of events can be gathered.

 \twocolumn

 \newpage
 \narrowtext
 \begin{table}[ht!]
 \begin{center}
 \begin{tabular}{ccc}
 & $\chi_{c0} + B^*$ & $\chi_{c0} + B^* + B_0$ \\
 $Br(B^- \to \pi^- \pi^- \pi^+)_{cut}$ & $1.18 \times
 10^{-7}$ & $1.06 \times 10^{-7}$  \\
 $Br(B^+ \to \pi^- \pi^+ \pi^+)_{cut}$ & $1.48 \times
 10^{-7}$ & $2.54 \times 10^{-7}$  \\
 ${\cal A}_{cut}$ & $0.11$ & $0.41$\\
 \end{tabular}
 \caption{Different contributions to the branching ratio and asymmetry in the
 decay channel $B^- \to \pi^- \pi^- \pi^+$. Both branching ratio and asymmetry
 are cut off according to the rules in (\ref{cut},\ref{cut-off})
 and $\sin \gamma = 0.82$.}
 \label{t:tab1}
 \end{center}
 \end{table}

 \end{document}